\documentclass[12pt]{article}

\begin{document}

\title{\addvspace{-10mm} {\normalsize \hfill CFNUL/99-02 }\\
\vskip-2mm
{\normalsize \hfill DFAQ/99/TH/03 } \\ 
\vskip-2mm
{\normalsize \hfill hep-ph/9908211}\\
\addvspace{15mm} Classical Nambu-Goldstone fields}
\author{Lu\'{\i}s Bento \\
{\normalsize \emph{Centro de F\'{\i}sica Nuclear da Universidade de Lisboa, }%
} \\
{\normalsize \emph{Avenida Prof. Gama Pinto 2, 1649-003 Lisboa, Portugal }} 
\\
\\
Zurab Berezhiani \\
{\normalsize \emph{INFN and Universit\`a di L'Aquila, I-67010 Coppito,
L'Aquila, Italy }} \\
{\normalsize \emph{Institute of Physics, Georgian Academy of Sciences,
380077 Tbilisi, Georgia }}}
\date{April, 2000}
\maketitle

\begin{abstract}
It is shown that a Nambu-Goldstone (NG) field may be coherently produced by
a large number of particles in spite of the fact that the NG bosons do not
couple to flavor conserving scalar densities like $\bar{\psi}\psi $. If a
flavor oscillation process takes place the phases of the pseudo-scalar or
flavor violating densities of different particles do not necessarily cancel
each other. The NG boson gets a macroscopic source whenever the total
(spontaneously broken) quantum number carried by the source particles
suffers a net increase or decrease in time. If the lepton numbers are
spontaneously broken such classical NG (majoron) fields may significantly
change the neutrino oscillation processes in stars pushing the observational
capabilities of neutrino-majoron couplings down to $m_{\nu }/300\,\mathrm{GeV%
}$.\newline
\strut \newline
\end{abstract}

\baselineskip=16pt


\newpage

As is well known, spontaneous violation of global symmetries leads to the
appearance of massless Nambu-Goldstone (NG) bosons in the particle
spectrum. The NG bosons not only have zero mass but they also have no
scalar potential terms such as $\phi ^{4}$. Nevertheless, they do not
mediate long range forces because they only have derivative couplings~\cite
{wein96}. 
This can be understood as follows. A NG boson $\phi $ associated with a
quantum number $\Lambda $ broken by the vacuum transforms under U(1)$%
_{\Lambda }$ as $\phi \rightarrow \phi +\alpha \,,$ $\alpha =\mathrm{const.}$%
, and all the other fields are made invariant in the unitary gauge.
The $\Lambda $ conservation law takes the form, ignoring for simplicity
possible mixing with other NG bosons, 
\begin{equation}
V_{\Lambda }\,\partial _{\mu }\partial ^{\mu }\phi +\partial _{\mu
}J_{\Lambda }^{\mu }=0\;,  \label{ddfidj}
\end{equation}
which is nothing but the equation of motion of the NG boson. $V_{\Lambda }$
is the scale of symmetry breaking and $J_{\Lambda }^{\mu }$ is the current
of the other fermion and boson particles. In other words, this equation
reads $\partial _{\mu }J^{\mu }=0$, where 
\begin{equation}
J^{\mu }=J_{\Lambda }^{\mu }+V_{\Lambda }\partial ^{\mu }\phi   \label{Jmu}
\end{equation}
is an exactly conserved current.

The source of a scalar field is of course a scalar density. The possible
fermion bilinears are the scalar and pseudo-scalar densities like $\overline{%
f}_{i}f_{j}$ and $\overline{f}_{i}\gamma _{5}f_{j}$ but they have to be
derived from the divergence of some current $J_{\Lambda }^{\mu }$. The
actual current depends on the particular theory and quantum number $\Lambda $%
. $J_{\Lambda }^{\mu }$ is determined in leading order by the quantum
numbers of the existing particles but receives also higher order corrections
with a more general flavor structure. In any case, it suffices to apply the
Dirac equation to arbitrary vector and axial-vector bilinears, written in
terms of the fermion mass eigenstates, to conclude, as follows from 
\begin{eqnarray}
&&\partial _{\mu }\,\overline{f}_{i}\gamma ^{\mu }\gamma
_{5}f_{j}=i(m_{i}+m_{j})\overline{f}_{i}\gamma _{5}f_{j}\ ,  \label{3} \\
&&\partial _{\mu }\,\overline{f}_{i}\gamma ^{\mu }f_{j}=i(m_{i}-m_{j})%
\overline{f}_{i}f_{j}\;,  \label{4}
\end{eqnarray}
that the NG bosons couple to pseudo-scalar densities (generally these can be
both flavor diagonal and non-diagonal) or to \emph{off-diagonal} scalar
densities. However couplings to \emph{flavor-diagonal} scalar densities are
not possible~\cite{chik81,gelm83}. The pseudo-scalars that are diagonal in
flavor vanish for free particle states and the flavor violating densities
depend on the relative phase of distinct flavors. The natural conclusion has
been that these relative phases cancel each other when summed over a large
number of particles and therefore long range '$1/r$' NG fields are not
possible. Only \emph{spin-dependent} '$1/r^{3}$' interaction potentials 
can exist~\cite{chik81,gelm83}.

However, there may be cases where certain quantum numbers are not conserved
in a very large scale. For example, the total lepton number is violated if
the neutrinos have Majorana masses, while the partial lepton numbers $L_{e}$%
, $L_{\mu }$ or $L_{\tau }$ are violated if they are also mixed (\emph{i.e.},
 if the non-diagonal elements of the neutrino mass matrix are non-zero in
the flavor basis). In other words, this means that in physical processes the
individual lepton numbers are violated by the neutrino oscillation
phenomena, e.g. in stars or other astrophysical objects. If the unconserved
quantum number $\Lambda $ is spontaneously broken, it implies the existence
of a massless NG boson -- majoron~\cite{chik81}.\footnote{%
We call such NG bosons majoron independently of whether the quantum number $%
\Lambda $ is the total lepton number $L$ or a partial lepton number $%
L_{e,\mu ,\tau }$.} As it was pointed out by one of us~\cite{bent97}, in
this case 
the non-vanishing divergency $\partial _{\mu }J_{\Lambda }^{\mu }$ generates
a classical field for the respective NG boson (majoron). The prototype of such
system is a flux of neutrinos undergoing a flavor oscillation process. The
key idea can be phrased as follows: as far as the neutrino oscillation
process implies a non-conservation of the currents like $J_{e}^{\mu }=\bar{%
\nu}_{e}\gamma ^{\mu }\nu _{e}$, etc.~associated with the partial lepton
numbers, then the conservation of the full current (\ref{Jmu}) should imply
the existence of a classical configuration of the corresponding majoron 
fields $\phi _{e}$, etc., with non-vanishing $\partial _{\mu }\phi _{e}$~\cite
{bent97}. (Clearly, a constant NG field does not carry much physical sense.)
This seems in contradiction with what was said about the interaction of a NG
boson with fermion scalar densities. It is the aim of this paper to show
that it is not so because the phases of wave functions of different flavors
are not independent from each other if transition processes take place
between them. In that event, the wave functions of such flavors interfere
constructively with each other and potentially form a macroscopic source of
the NG boson built out of non-diagonal densities like $\overline{f}%
_{i}\gamma _{5}f_{j}$ or $\overline{f}_{i}f_{j}$, in the basis of the mass
eigenstates. To see this one has to be more specific about the nature of the
fermion current and equations of motion. \strut 

Consider a NG boson associated with the spontaneous breaking of a partial
lepton number $\Lambda $, which could be $L_{e}$, $L_{\mu }$, $L_{\tau }$,
or any combination of these. Denoting the neutrinos 
by $f^{a}$, with quantum numbers $\Lambda _{a}$, and restricting ourselves
to the fermion current, the equation of motion (\ref{ddfidj}) of the
corresponding NG boson (majoron) is given in leading order by 
\begin{equation}
\partial _{\mu }\partial ^{\mu }\phi =-\frac{1}{V_{\Lambda }}%
\sum_{a}\partial _{\mu }\left( \overline{f}^{a}\gamma ^{\mu }\Lambda
_{a}f^{a}\right) \;.  \label{ddfiff}
\end{equation}
After applying the fermion equations of motion, the interactions that
violate the lepton number $\Lambda $ emerge in the second term. They vary
from model to model but include in general neutrino Majorana masses. These
are the lowest dimension terms that violate $\Lambda $ and we confine to
them here neglecting any higher order flavor violating gauge or Yukawa
interactions. For simplicity, the neutrino mass matrix, $m$, is assumed to
be real in the weak basis. Denoting by $\nu _{L}^{a}$, $\nu _{L}^{aC}$ the
left-handed neutrino fields and their charge conjugates, the equations of
motion are of the form 
\begin{equation}
i\,\partial \hspace{-0.22cm}/\,\nu _{L}^{a}=m_{ab}\,\nu _{L}^{aC}+V_{\mu
}^{a}\,\gamma ^{\mu }\,\nu _{L}^{a}\;.  \label{dna}
\end{equation}
The potentials $V_{\mu }^{a}$ account for the \textit{local} neutral and
(Fierz-transformed) charge current interactions with the medium~\cite{kuo89}
and couplings to the NG boson namely, 
\begin{equation}
{\mathcal{L}}_{\nu \nu \phi }=\frac{1}{V_{\Lambda }}\partial _{\mu }\phi \;\,%
\overline{\nu _{L}^{a}}\,\gamma ^{\mu }\Lambda _{a}\,\nu _{L}^{a}\;.
\label{lnnf}
\end{equation}
The result is 
\begin{equation}
\partial _{\mu }\partial ^{\mu }\phi =\frac{i}{V_{\Lambda }}\left[ \overline{%
\nu _{L}^{a}}\,\left( \Lambda \,m\right) _{ab}\,\nu _{L}^{bC}-\overline{\nu
_{L}^{aC}}\,\left( m\,\Lambda \right) _{ab}\,\nu _{L}^{b}\right] \;,
\label{ddfiop}
\end{equation}
where the quantum number $\Lambda $ is written in matrix form. For example,
one can consider the anomaly free quantum number $L_{e}-L_{\mu }$, in which
case the (2-flavor) matrices of leptonic charge and neutrino mass have the
form 
\begin{equation}
\Lambda =\left( 
\begin{array}{cc}
{1} & {0} \\ 
{0} & {-1}
\end{array}
\right) ,~~~~~m=\left( 
\begin{array}{cc}
{M_{1}} & {M} \\ 
{M} & {M_{2}}
\end{array}
\right) ,  \label{matr}
\end{equation}
respectively, where $M$ is a $\Lambda =L_{e}-L_{\mu }$ conserving entry and $%
M_{1},M_{2}$ are the $\Lambda $-violating ones. The result above illustrates
also that in scattering processes the effective coupling constants of
neutrinos to majorons are essentially the neutrino masses divided by the
scale of global symmetry breaking, $V_{\Lambda }$. This point will be called
later when discussing the observational implications.

In order to evaluate the source terms over a system of particles one needs
to relate the field operators with the wave functions. The chiral fields $%
\nu _{L}$, $\nu _{L}^{C}$\ are the left and right-handed projections of the
Majorana fields 
\begin{equation}
\nu _{L}+\nu _{L}^{C}=\int a\,\psi +a^{\dagger }\,\psi ^{C}\;,  \label{chi}
\end{equation}
where $a$, $a^{\dagger }$ are annihilation and creation operators and $\psi $
single particle wave functions. The expectation value of the operators in
Eq.~(\ref{ddfiop}) gives then a sum over all the existing $\nu $ particles
in terms of their wave functions ($\gamma _{5}$ is equal to $-1$ for
left-handed spinors): 
\begin{equation}
\partial _{\mu }\partial ^{\mu }\phi =\frac{i}{V_{\Lambda }}\sum_{\nu }%
\overline{\psi ^{a}}\,\left( m\Lambda +\Lambda m\right) _{ab}\gamma
_{5}\,\psi ^{b}\;.  \label{ddfiwf}
\end{equation}
The second member contains precisely the pseudo-scalar densities present in
Eq.~(\ref{3}). It remains to establish the $\psi $ equations of motion. They
are non-linear at the operator level due to Majorana mass terms but the wave
functions obey linear equations as follows from Eqs.~(\ref{dna}) and (\ref
{chi}): 
\begin{equation}
i\,\partial \hspace{-0.22cm}/\psi ^{a}=m_{ab}\,\psi ^{b}-V_{\mu
}^{a}\,\gamma ^{\mu }\,\gamma _{5}\psi ^{a}\;.  \label{dpsi}
\end{equation}

These are the equations of motion relevant for the most common cases of
neutrino propagation in matter or vacuum. What is usually done is to
separate the spin and flavor degrees of freedom by expressing the wave
function as a product of a left-handed spinor $\psi _{0}^{\alpha }$ ($\gamma
_{5}\psi _{0}=-\psi _{0}$), solution of the zero mass Dirac equation 
\begin{equation}
i\,\partial \hspace{-0.22cm}/\psi _{0}=0\;,  \label{dpsi0}
\end{equation}
and a flavor-valued wave function $\varphi ^{a}$ that obeys a well known
evolution equation as a function of the distance travelled by each neutrino.
Here, one has to go beyond that approximation because the pseudo-scalar
densities in Eq.~(\ref{ddfiwf}) vanish for spinors with well defined
chirality. An approximate solution of Eq.~(\ref{dpsi}) is 
\begin{equation}
\psi ^{a\alpha }(x)\cong \psi _{0}^{\alpha }\,\varphi ^{a}+\gamma _{\alpha
\beta }^{0}\frac{m_{ab}}{2E}\,\psi _{0}^{\beta }\,\varphi ^{b}\;,
\label{psiaa}
\end{equation}
which applies for the most common cases where only the scalar components $%
V_{0}^{a}$ exist and even for vector potentials $\vec{V}^{a}$ that are
parallel to the neutrino velocity $\vec{v}$. The wave function $\varphi
^{a}\ $is considered as a function of the distance travelled by the neutrino
to the point $\vec{x}$, $s=\vec{x}\cdot \!\vec{v}-\vec{x}_{0}\cdot \!\vec{v}$%
, while the dependence on the variable $t-s$ (constant along the neutrino
trajectory) is totally absorbed in $\psi _{0}$. $\varphi ^{a}\ $obeys the
familiar evolution equation~\cite{kuo89,wolf78,mikh86} 
\begin{equation}
i\frac{d\varphi ^{a}}{ds}=\left( \frac{m^{2}}{2E}+V_{\mu }\,v^{\mu }\right)
_{ab}\varphi ^{b}\;,  \label{dfids}
\end{equation}
where $v^{\mu }=p^{\mu }\!/\!E$ is the neutrino velocity and $V_{\mu
}^{ab}=\delta _{ab}\,V_{\mu }^{a}\,$.

Using these wave functions, the majoron equation of motion (\ref{ddfiwf})
reads in leading order 
\begin{equation}
\partial _{\mu }\partial ^{\mu }\phi =\frac{i}{V_{\Lambda }}\sum_{\nu
}\varphi ^{a\dagger }\left[ \Lambda ,\frac{m^{2}}{2E}\right] _{ab}\!\varphi
^{b}\,\psi _{0}^{\dagger }\psi _{0}\;.  \label{ddfif1}
\end{equation}
It is now clear that no flavor conserving terms contribute to the majoron
source either in the weak basis where $\Lambda $ is diagonal or in the mass
eigenstate basis where $m$ is a diagonal matrix. The existence of a majoron
source relies on the interference between different flavor components of the
neutrino wave functions. But that simply means the existence of neutrino
oscillations. Another necessary condition is that the flavor dynamics
Hamiltonian does not conserve the quantum number $\Lambda $. Here, as a
result of our particular specification of the neutrino interactions, $%
\Lambda $ is only violated by the mass matrix. In general there could be
also some additional non-standard neutrino flavor-violating interactions
with matter constituents whose effects can be comprised in the potential $%
V_{\mu }$ in Eq.~(\ref{dfids}). The remarkable feature is the absence of
cancellation due to the arbitrariness of the wave function initial phases:
each neutrino contributes with a term that only depends on the phase
invariants $\psi _{0}^{\dagger }\psi _{0}$ and $\varphi ^{a\dagger }\varphi
^{b}$. This crucial fact permits that in certain circumstances the
contributions from a large number of particles add to each other with a
definite sign making so a source of macroscopic dimensions. The precise
understanding of the nature of those circumstances is the subject of next
discussion.

Making use of Eq.~(\ref{dfids}) one can write the last equation as 
\begin{equation}
\partial _{\mu }\partial ^{\mu }\phi =-\frac{1}{V_{\Lambda }}\sum_{\nu }\psi
_{0}^{\dagger }\psi _{0}\,\frac{d}{ds}\varphi ^{a\dagger }\Lambda
_{a}\varphi ^{a}\;.  \label{ddfif2}
\end{equation}
Again, the second member does not depend on the initial phases of the
individual particles. It rather depends on whether there is a net increase
or decrease of the $\Lambda $ number carried by the neutrinos. Take the
example of a large system where only $\nu _{e}$ are produced out of
electrons captured in nuclear reactions. The oscillations $\nu
_{e}\rightarrow \nu _{\mu }$ necessarily lead to a total decrease of the
partial lepton number $\Lambda =L_{e}-L_{\mu }$ that was initially carried
by electrons. If $\Lambda $ is a spontaneously broken quantum number, then
the equation of motion of the corresponding NG boson $\phi _{\Lambda }$ is
(we ignore a possible mixing with other NG bosons~\cite{bent97}): 
\begin{equation}
\partial _{\mu }\partial ^{\mu }\phi _{\Lambda }=-\frac{1}{V_{\Lambda }}%
\sum_{\nu }\psi ^{\dagger }\psi \,\frac{d(P_{\nu _{e}}-P_{\nu _{\mu }})}{ds}%
=-\frac{2}{V_{\Lambda }}\sum_{\nu }\psi ^{\dagger }\psi \,\frac{dP_{\nu _{e}}%
}{ds}\;,  \label{ddfipe}
\end{equation}
where for simplicity we have assumed that only $\nu _{e}\leftrightarrow \nu
_{\mu }$ takes place and thus $P_{\nu _{e}}+P_{\nu _{\mu }}=1$. Therefore,
the shape of the majoron field $\phi _{\Lambda }$ is determined in terms of
the probability $P_{\nu _{e}}(s)=\varphi ^{e\dagger }\varphi
^{e}\!/\!\varphi ^{\dagger }\varphi $ of observing the $\nu _{e}$ flavor.
Since in the presence of oscillations this probability is essentially
smaller than $P_{\nu _{e}}(0)=1$, the total $\phi _{\Lambda }$ source charge
is positive. The consequence is a long range $\phi _{\Lambda }$ classical
field that propagates at the velocity of light. The $\phi _{\Lambda }$
sources exist where $L_{e}-L_{\mu }$ is either created or destroyed and the
size of these regions is determined by the neutrino oscillation length, or
resonance length in the case of resonant oscillations in a medium.

An explicit example is the well known case of two flavor oscillations in
vacuum~\cite{kuo89}. With a mixing angle $\theta $ and $\Delta m^{2}$ mass
difference the $\nu _{e}$ flavor probability is 
\begin{equation}
P_{\nu _{e}}(s)=1-\sin ^{2}2\theta \,\sin ^{2}\left( \frac{\Delta m^{2}}{4E}%
s\right) \;,  \label{pe}
\end{equation}
as a function of the distance $s$ travelled by the neutrino from its
production point. The equation of motion of the NG field becomes 
\begin{equation}
\partial _{\mu }\partial ^{\mu }\phi _{\Lambda }=\frac{2}{V_{e}}\sum_{\nu
}\psi ^{\dagger }\psi \,\sin ^{2}2\theta \,\frac{\Delta m^{2}}{4E}\,\sin
\left( \frac{\Delta m^{2}}{4E}s\right) \;.  \label{ddfipe2}
\end{equation}
The source term changes sign as a function of $s$ but the scale of spatial
variation is the flavor oscillation wavelength, $4E/\Delta m^{2}$, not any
scale related to the number density or linear momentum of the particles. The
neutrinos produced too far away give a vanishing contribution because the
coherence is lost, but the ones produced within a sphere of the scale of the
oscillation wavelength give a non-zero source term. Its sign depends on the
distance to the point where each $\nu _{e}$ was produced but, if one
integrates over all neutrinos, the result is finite as shows the following
example. If the reactor or star produces $\nu _{e}$ neutrinos in a
stationary basis, the NG field obeys a Poisson equation and the total
'charge', the volume integral of the second member, is positive. This can be
seen from Eqs.~(\ref{ddfipe}) and (\ref{pe}): the volume integration is
trivial if one considers neutrinos propagating in plane waves and the
average value of the probability $P_{\nu _{e}}$ over the neutrino spectrum
converges to $1-\sin ^{2}2\theta /2$ at distances much larger than the
oscillation length. For a total $\nu $ luminosity equal to $\dot{N}_{\nu },$
the total 'charge' is $2/V_{\Lambda }$ times the number of neutrinos that
undergo the transition $\nu _{e}\rightarrow \nu _{\mu }$ per unity of time, 
\begin{equation}
\frac{2}{V_{\Lambda }}\frac{dN}{dt}(\nu _{e}\rightarrow \nu _{\mu })=\frac{2%
}{V_{\Lambda }}\dot{N}_{\nu }\left( 1-\left\langle P_{\nu _{e}}(\infty
)\right\rangle \right) =\frac{1}{V_{\Lambda }}\dot{N}_{\nu }\sin ^{2}2\theta
\;.  \label{tcharge}
\end{equation}
This quantity has a definite sign, regardless of the size of the reactor,
the only assumption is the nuclear reactions feed the system with $\nu _{e}$
at a constant rate in time. That means that the NG field is just like the
electrostatic potential due to an electric charge distribution: it exits at
distances much larger than the size of the source (the place where neutrinos
oscillate) with a Coulombian '$1/r$' shape. What is really different is the
nature of the charge. The 'charge' of this NG field is the time rate of
decreasing of the total $L_{e}-L_{\mu }$ number carried by leptons.

\strut In the case of neutrino propagation in the medium, the neutrino
oscillation is typically suppressed by the effects of the coherent
interaction with the matter constituents. However, there can be a resonant
neutrino conversion, so called MSW (Mikheyev-Smirnov-Wolfenstein)
oscillation~\cite{wolf78,mikh86}, where almost complete $\nu _{e}\rightarrow
\nu _{\mu }$ conversion takes place within the length of the resonance area.
Then, the source of the NG field lies in the resonance shell, spheric in the
case of a star, and the total NG charge is finite (provided that the $\nu
_{e}\rightarrow \nu _{\mu }$ conversions are not  compensated by an exactly
equal number of $\bar{\nu}_{e}\rightarrow \bar{\nu}_{\mu }$ or $\nu _{\mu
}\rightarrow \nu _{e}$ conversions, which might happen if $\bar{\nu}_{e}$ or 
$\nu _{\mu }$ are also produced in the reactor or star). Consequently, a
Coulombian field is produced with
\begin{equation}
\phi _{\Lambda }=\frac{2}{V_{\Lambda }}\frac{1}{4\pi r}\frac{dN}{dt}(\nu
_{e}\rightarrow \nu _{\mu })
\end{equation}
at radius $r$ larger than the radius of the resonance shell. The field is
constant in the region inside that shell.

The fermions interact with the long range NG field through its gradient. As
shows Eq.~(\ref{lnnf} ), $\partial _{\mu }\phi _{\Lambda }$ acts as a vector
potential on the neutrinos (and charged leptons). In the case illustrated
above a radially moving neutrino 'feels' a potential energy equal to
\begin{equation}
V_{NG}=\mp \frac{1}{V_{\Lambda }}\vec{v}_{\nu }\!\cdot \!\vec{\nabla}\phi
_{\Lambda }=\pm \frac{2}{V_{\Lambda }^{2}}\frac{1}{4\pi r^{2}}\frac{dN}{dt}%
(\nu _{e}\rightarrow \nu _{\mu })\;,
\end{equation}
where the upper sign is for $\nu _{e}$, $\bar{\nu}_{\mu }$\ and the lower
sign for $\nu _{\mu }$, $\bar{\nu}_{e}$. The derivative nature of the NG
boson couplings gives an effective $\nu \nu \phi $ coupling constant equal
to $V_{\Lambda }^{-1}$ over the distance to the $\phi _{\Lambda }$ source.
This is typically an extremely small number. However neutrino oscillations
are sensitive to very tiny external potentials. In addition the NG boson
couplings depend on the particle quantum numbers and such a non-universality
makes them potentially important for the neutrino oscillations themselves
(see Eq.~(\ref{dfids})). Because these are long-range fields one can
conceive that a NG field generated in some region of a star may affect the
propagation of other neutrinos or flavors outside that region or even out of
the star.

Some possible effects can be devised concerning the Supernova neutrinos~\cite
{bent97}. The NG potentials are proportional to the $\nu $ luminosity over $%
(rV_{\Lambda })^{2}$. In the case of Supernovae this number is comparable to
the values of $\Delta m^{2}/2E$ that are interesting for the solar $\nu $
solutions for scales $V_{\Lambda }$ as large as the weak breaking scale. It
was shown~\cite{bent97} that the NG fields may then have a significant role
in \emph{resonant} $\nu $ oscillations. A special feature is the dependence
of the effects on the absolute magnitude of the $\nu $ fluxes. The signature
is a surprise, \emph{i.e.} oscillation patterns of Supernova neutrinos that
are in contradiction with the solar, atmospheric and laboratory neutrinos
observations and/or that turn off as the $\nu $ fluxes decay after the first
instants of Supernova $\nu $ emission. This may be observed with detectors
capable of detecting Supernova~ neutrinos and measuring their energy spectra
or time evolution. Then, one could go beyond present experimental and
astrophysical limits on the majoron effective couplings to neutrinos~\cite
{SN} (these are typically of the order of the neutrino masses over the scale 
$V_{\Lambda }$, see Eq.~(\ref{ddfiop})), provided that the scale of symmetry
breaking is under 1 TeV~\cite{ABST}. The NG field could have some impact on
solar neutrinos too if the scale $V_{\Lambda }$ is below KeV. Though such a
situation is not very plausible, there exist some models~\cite{BSV} in which 
$V_{\Lambda }$ can be very low and thus the effective majoron-neutrino
coupling constants rather large, up to order $10^{-2}-10^{-3}$, still
evading the present limits.

\strut The following remark is in order. The physical features of generating
the classical majoron field due to neutrino oscillation is very much
different from the mechanism of the majoron production due to the neutrino
coherent scattering on the matter components of a medium -- so called matter
induced neutrino decay with majoron emission~\cite{majdec}. In the latter
case majorons are produced as particles since the medium provides an energy
splitting between the neutrino and antineutrino states and so the
transitions either $\nu \rightarrow \bar{\nu}+\phi $ or $\bar{\nu}%
\rightarrow \nu +\phi $ are possible, depending on the neutrino flavor and
on the chemical content of the medium (\emph{i.e.}, which of two states
becomes ''heavier'' in the matter background). 
Very low-momentum majorons could also be produced as a result of 
(stimulated) neutrino decays in the Early Universe~\cite{madsen92},
forming a sort of Bose condensate 
with large occupation numbers in the low momemtum part of the spectrum, 
but still made of individual majoron particles.
On the opposite, the
production of the classical majoron field rather resembles the situation
with classical electromagnetic field produced by the electric current.

Concluding, we have shown that if a process of flavor oscillation
takes place a pseudo-scalar density like the one of Eq.~(\ref{ddfiwf}) may
give rise to a long-range NG field as a result of the constructive
interference between the wave functions of different mass eigenstates.
Furthermore, the source of the NG boson is the time rate of decreasing of
the quantum number associated with it, which is nothing but the term $%
-\partial _{\mu }J_{\Lambda }^{\mu }$ in Eq.~(\ref{ddfidj})~\cite{bent97}.
If the lepton numbers are spontaneously broken at the Fermi energy scale or
below, and that means effective neutrino-majoron coupling constants as low
as $m_{\nu }/300\,\mathrm{GeV}$, the associated majoron fields are still
significant enough for neutrino oscillations in Supernovae whose spectra may
then show evidence of their existence beyond the limits~\cite{SN} that can
be reached from laboratory or astrophysical scattering processes.

\section*{Acknowledgments}

We would like to thank Sasha Dolgov for useful discussions. L. B.
acknowledges the support of FCT under the grant PESO/P/PRO/1250/98.

\strut

\end{document}